\documentclass[12pt, a4paper]{article}

\usepackage{physics}
\usepackage{enumerate}
\usepackage{graphicx}
\usepackage{array}
\usepackage{makecell}
\usepackage{enumitem}
\usepackage{pgfplots}
\usepackage{hyperref}
\usepackage{booktabs}
\usepackage{chemformula}
\usepackage{subcaption}
\usepackage{pgfplotstable}
\usepackage{tikz,pgfplots}
\usepackage{amsmath}  

\AtBeginDocument{\RenewCommandCopy\qty\SI} 
\usepackage{geometry}
\pgfplotsset{compat=1.14}

\usepackage[backend=biber,style=chem-acs]{biblatex}
\begin{document}

\begin{center}
\Large {\textbf {Singlet-Triplet Kondo Effect in Blatter Radical Molecular Junctions: Zero-bias Anomalies and Magnetoresistance }}
\end{center}

\begin{center}

\small{Gautam Mitra$^{1,2}$, Jueting Zheng$^{1,3}$, Karen Schaefer$^4$, Michael Deffner$^{4,5}$, Jonathan Z. Low$^6$, Luis M. Campos$^6$, Carmen Herrmann$^{4,5}$, Theo A. Costi$^7$ and Elke Scheer$^1$}\\

\hfill

$^1$ Department of Physics, University of Konstanz, 78464 Konstanz, Germany\\
$^2$ Institute of Condensed Matter and Nanosciences, Université Catholique de Louvain (UCLouvain), 1348 Louvain-la-Neuve, Belgium\\
$^3$ State Key Laboratory of Physical Chemistry of Solid Surfaces, College of Chemistry
and Chemical Engineering, Xiamen University, 361005 Xiamen, China\\
$^4$ Department of Chemistry, University of Hamburg, 22761 Hamburg, Germany\\
$^5$The Hamburg Centre for Ultrafast Imaging, Hamburg 22761, Germany\\
$^6$ Department of Chemistry, Columbia University,  New York, NY 10027, USA\\
$^7$ Peter Grünberg Institut, Forschungszentrum Jülich, 52425 Jülich, Germany\\

\end{center}

\hfill

\begin{abstract}
\noindent The Blatter radical has been suggested as a building block in future molecular spintronic devices due to its radical character and expected long-spin lifetime. However, whether and how the radical character manifests itself
in the charge transport and magnetotransport properties seems to depend on the environment or has not yet been studied. 
 Here, we investigate single-molecule junctions of the Blatter radical molecule in a mechanically controlled break junction device at low temperature. Differential conductance spectroscopy on individual junctions shows two types of zero-bias anomalies attributed to Kondo resonances revealing the ability to retain the open-shell nature of the radical molecule in a two-terminal device. Additionally, a high negative magnetoresistance is also observed in junctions without showing a zero-bias peak. We posit that the high negative magnetoresistance is due to the effect of a singlet-triplet Kondo effect under magnetic field originating from a double-quantum-dot system consisting of a Blatter radical molecule with a strong correlation to a second side-coupled molecule. Our findings not only provide the possibility of using the Blatter radical in a two-terminal system under cryogenic conditions but also reveal the magnetotransport properties emerging from different configurations of the molecule inside a junction.

\end{abstract}

\section*{Introduction}

Stable organic radical molecules are emerging to become key components for spin-based molecular electronic materials and devices \cite{Ji2020b,Wang2017,Deumal2021,Tomlinson2014,Ratera2012, Sanvito2011a, Herrmann2010a, Mas-Torrent2012}. They are open-shell systems which do not readily decompose or chemically react 
under standard conditions and whose stability can be engineered by virtue of chemical design \cite{Shu2023,Tang2020,Grillo2012,Liu2013,Mullegger2013,Hicks2007}. The 1,2,4-benzotriazin-4-yl radical commonly known as the Blatter radical is one of such species, first reported in 1968 \cite{Blatter1968}, gaining interest owing to its many exciting physical and chemical properties \cite{Miura2015,Yan2011,Berezin2013,Constantinides2011}.  The stability of the Blatter radical under various circumstances enables to incorporate it into manifold electronic devices \cite{Ciccullo2016, Low2019a, Patera2019}, controlled polymerization setups\cite{Areephong2016}, photodetectors \cite{Zheng2015} and thermoelectric devices\cite{Hurtado-Gallego2022}. In addition, the presence of an unpaired electron in this  molecule can lead to interesting magnetic properties,  
when interfaced with metallic substrates and electrodes \cite{Low2019a, Patera2019, Khurana2020,Ji2020}.\\

\begin{figure}[t]
    \centering
    \includegraphics[width = 1\textwidth]{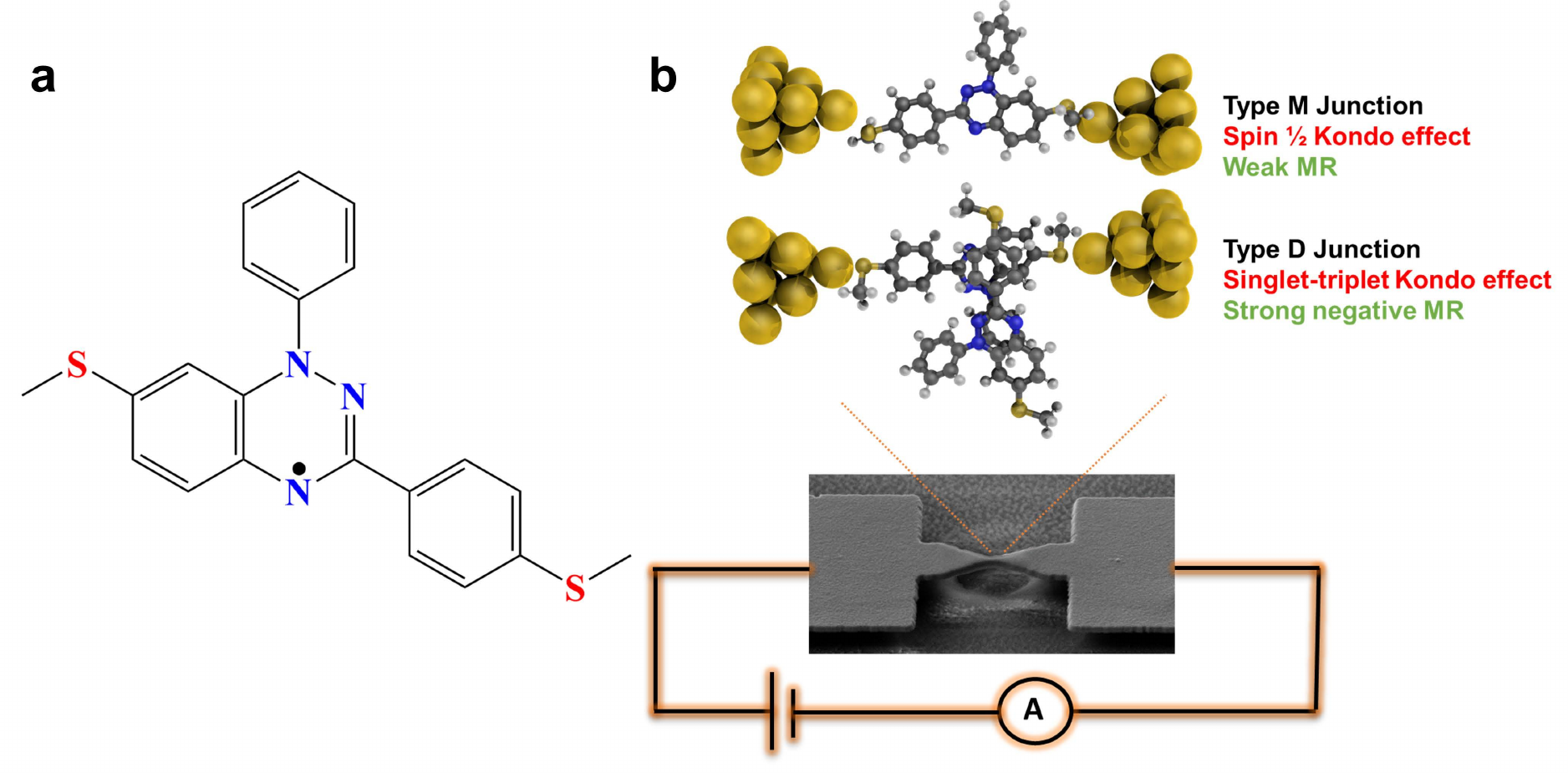}
    \caption{(a) The chemical structure of the Blatter radical molecule used in this study.  (b) Illustration of the types of Blatter radical molecule junction formed between two Au electrodes on an MCBJ device. Type M indicates a monomer junction and Type D indicates a dimer junction.}
    \label{fig:mitra_blatter_fig_1}
\end{figure}

Blatter radicals typically have the unpaired electron delocalized on the molecule (Figure \ref{fig:mitra_blatter_fig_1}a). This spatial delocalization of the unpaired electron gives rise to the exceptionally high stability of the radical. Recently, these systems have been subject to studies under different environments of metal-molecule interfaces and have shown to retain the open-shell nature in thin films \cite{Ciccullo2016}, and upon adsorption on the Au(111) surface under high vacuum \cite{Low2019a}. Typically, Kondo spectroscopy is used as an evidence to interrogate the  molecular spin of the radical \cite{Madhavan1998,Ternes2009}. Previously, using scanning tunneling microscopy (STM) techniques under ultrahigh vacuum conditions, a Kondo resonance arising from the interaction between the unpaired electron of the Blatter radical with the conduction electrons was revealed by Repp and coworkers \cite{Patera2019}. The singly occupied molecular orbital (SOMO) of the Blatter radical lies close to the Fermi level of Au and was suggested to be very sensitive to the environment around the molecule \cite{Low2019a}. Little emphasis has been given to the study of Blatter junctions in a two-terminal metal-molecule-metal system mainly due to the suspected loss of the open-shell character upon oxidation, as reported in solution-based STM-break junction experiments \cite{Low2019a}. Later, this issue was tackled by functionalizing the radical by adding an electron-withdrawing group, which lowers the energy levels inside the molecule giving more stability and preventing oxidation \cite{Hurtado-Gallego2022}. Recently, Jiang et al. explored this problem theoretically using first-principles quantum transport calculations \cite{Jiang2023}. Their calculations suggest that the Blatter radical can successfully retain its open-shell character in a junction between Au electrodes irrespective of the environmental conditions since the dative bonding at the molecule electrode interface does not alter the spin states of the radical. 
\\

We posit that stable organic radical junctions can also manifest unconventional magnetotransport behavior in combination with spectroscopic evidence for the Kondo effect and that both can be tuned by the coupling strength to the electrodes. To the best of our knowledge there are no experimental magnetoresistance (MR) studies to date that focus on radicals with a delocalized unpaired electron in the current pathway using \emph{in-situ} tunable, yet sufficiently stable electrodes. To better understand the intrinsic properties of open shell systems, here we carry out the first systematic transport study of the Blatter radical in a mechanically controlled break junction (MCBJ) operated at low temperature. We were able to produce Blatter molecule junctions with increased stability in a two-terminal device and the signature of the open-shell nature of the radical is confirmed using $dI/dV$ spectroscopy.  The Blatter radical was synthesized with two methylthioether units (--SCH$_3$) to enable coupling with the Au electrodes \cite{Low2019a}. We have deposited the molecule by using a dropcasting technique and carried out the electronic transport measurements of the individual junctions in cryogenic vacuum at a temperature of 4.2 K and under magnetic fields. Kondo resonances and magnetoresistance (MR) originating from stable organic radical molecule junctions have been studied with the MCBJ technique before \cite{Low2019a, Frisenda2015c, Hayakawa2016,Zhang2024}. But in all previous cases, the electronic lone-pair was located in a specific site of the molecule and was sterically protected.  We have found two types of Kondo resonances, presumably originating from  different configurations of the individual junctions of the Blatter radical molecule. One type of the observed Kondo resonances features a Kondo temperature ($T_{\rm K}$) of $30 - 35$ K with a broad and asymmetric line shape and the other type has a $T_{\rm K}$ of $11 - 17$ K with a more symmetric and narrow line shape. We have confirmed the Kondo nature with the evolution of the zero bias peak as a function of the magnetic field. These exhibit only a very weak negative finite-bias MR, as predicted by the usual Kondo effect at finite-bias (see Figure S1 in SI). However, we have also observed a significant negative finite-bias MR in certain junctions without having a zero-bias peak at the measurement temperature of 4.2 K. We find a maximum change of 21\% in resistance with an applied magnetic field of $\pm$ 8 T and a subsequent saturation of MR for higher fields. These results highlight the stability of the unpaired electron in the Blatter radical molecule junctions in low temperature measurements with a MCBJ device. Finally, the negative MR at finite bias is explained in the framework of a singlet-triplet Kondo model. We have illustrated that the presence of an extra side-coupled molecule in the Blatter junction with an antiferromagnetic exchange interaction between the unpaired spins on the two radicals results in a possible scenario of negative MR at finite bias voltage and at a temperature similar to the experiment. A summary of our results from different types of junction formation is schematically shown in Figure \ref{fig:mitra_blatter_fig_1}b.

\section*{Results and Discussions}

\subsection*{Statistical investigations}

The Blatter radical molecule with two methylthioether units was synthesized and characterized as reported earlier \cite{Low2019a}. The freshly prepared solution of the molecule is prepared in tetrahydrofuran (THF) and dropcast onto a gold (Au) MCBJ sample at room temperature in a nitrogen atmosphere and properly dried for 45 min. The sample is then characterized using a dipstick equipped with a MCBJ mechanism pumped to high vacuum and cooled down to 4.2 K inside a liquid-He bath cryostat. More details of sample preparation and measurement techniques are described in the Methods section.\\

\begin{figure}[!b]
    \centering
    \includegraphics[width = 1\textwidth]{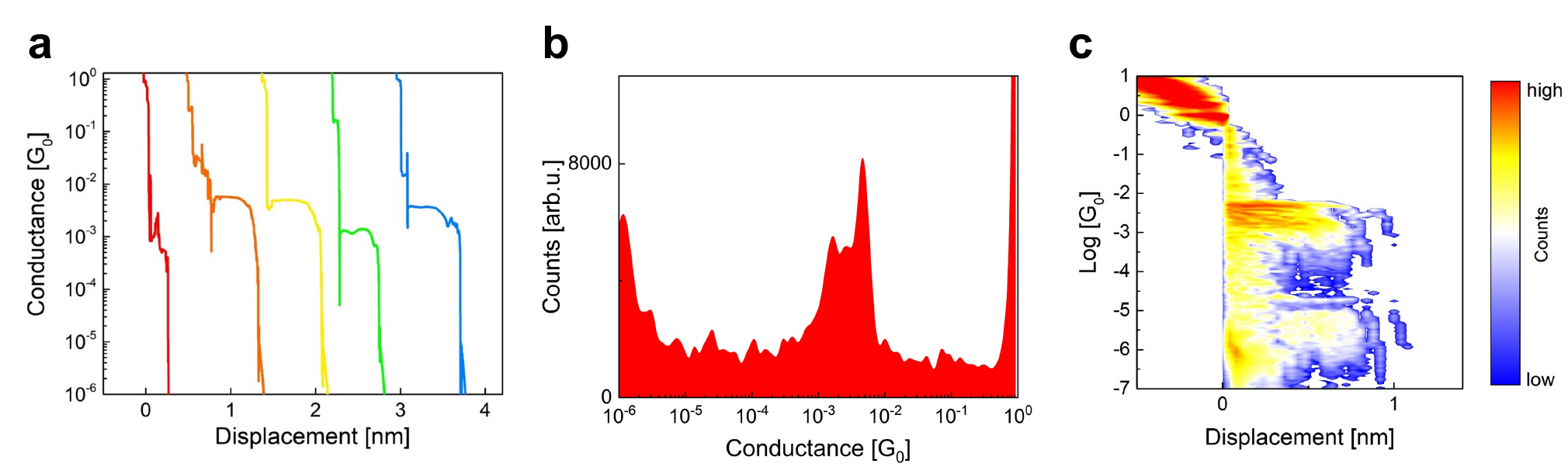}
    \caption{(a) Examples of typical opening traces of Blatter radical molecule junctions using the MCBJ method at 4.2 K. (b) 1D conductance histogram built from 595 opening traces measured at 4.2 K with 100 mV bias. (c) Two-dimensional histogram calculated from the same data as shown in (b) }
    \label{fig:histo}
\end{figure}

First, we measure and analyze the opening traces of the Blatter radical molecule with an applied bias of 100 mV at 4.2 K. Figure \ref{fig:histo}a shows typical examples of these opening traces as a function of electrode displacement. Au single-atom contacts with 1$G_{0}$ are observed in all the opening curves, where $G_{0} = 2e^2/h$ is the conductance quantum. Below 1$G_{0}$, upon further stretching of the electrodes, conductance plateaus are formed corresponding to the single-molecule junctions of the Blatter radical molecule bridged using methylthioether anchoring units. These plateaus shows a maximum displacement up to 0.6 nm before breaking down to the tunneling regime, with conductance $G$ below $10^{-6}G_{0}$. In Figure \ref{fig:histo}b, we have constructed a conductance histogram with logarithmic binning of 595 opening traces without any data selection. We observed major conductance peaks between $10^{-2}$ and $10^{-3}G_{0}$. We attribute multiple peaks observed in this conductance range to different configurations of molecule junctions formed between the Au electrodes. Flat plateaus observed in the opening traces correspond to fully stretched junctions of individual molecules whilst partially connected junctions show fluctuations in conductance as the electrode displacement is increased. This behavior is also clearly observed in the two-dimensional conductance-distance histogram as shown in Figure \ref{fig:histo}c. We note that there is also the possibility of dimers with two parallel or side-coupled molecules which are supported by our quantum chemical calculations using density functional theory (DFT, see the Supporting Information (SI)), the main results of which can be summarized as follows: 
First, the  Blatter radicals investigated here bind substantially to each other, as reported earlier for similar species \cite{kert18, deum21}. Second, the simulated zero-bias conductance values are in the range of $10^{-2}$ to $10^{-3}$ G$_0$, with the monomers (denoted as M) at the lower end of this range (in line with \cite{Jiang2023}, see SI). These values agree well with the experimentally observed range. Knowing that DFT has the tendency to overestimate the conductance by up to around an order of magnitude~\cite{Feng2016,Strange2011,Koentopp2008,Ke2007}, it appears likely that the structures giving rise to higher conductance values in the simulations, hence dimers (denoted as D), correspond to the ones realized in the experiments. This interpretation is in line with the relatively strong binding between the radicals. We will come back to this aspect further below.\\

The observed conductance peak in the histogram is also in close agreement with the previous studies of the same molecule on a solution-based scanning electron microscopy break junction (STM-BJ) experiment \cite{Low2019a} and STM studies on thin films of the Blatter radical on a Au surface at room temperature \cite{Patera2019}. The small conductance obtained from the Blatter radical molecule junctions were very recently tackled theoretically by Jiang et al. using first-principles quantum transport calculations \cite{Jiang2023}. They attribute their unexpectedly low conductance to singly occupied frontier orbitals (SOMO and SUMO, respectively) being weakly electronically coupled to the Au electrodes. Hence, the Blatter radical may retain its unpaired spin upon contacting with Au electrodes in more instances than previously thought. In order to verify this hypothesis, we have carried out several magnetotransport measurements on individual junctions as explained below. \\

\subsection*{Transport spectroscopy in individual junctions: Kondo resonances}

\begin{figure}[t]
    \centering
    \includegraphics[width = 1\textwidth]{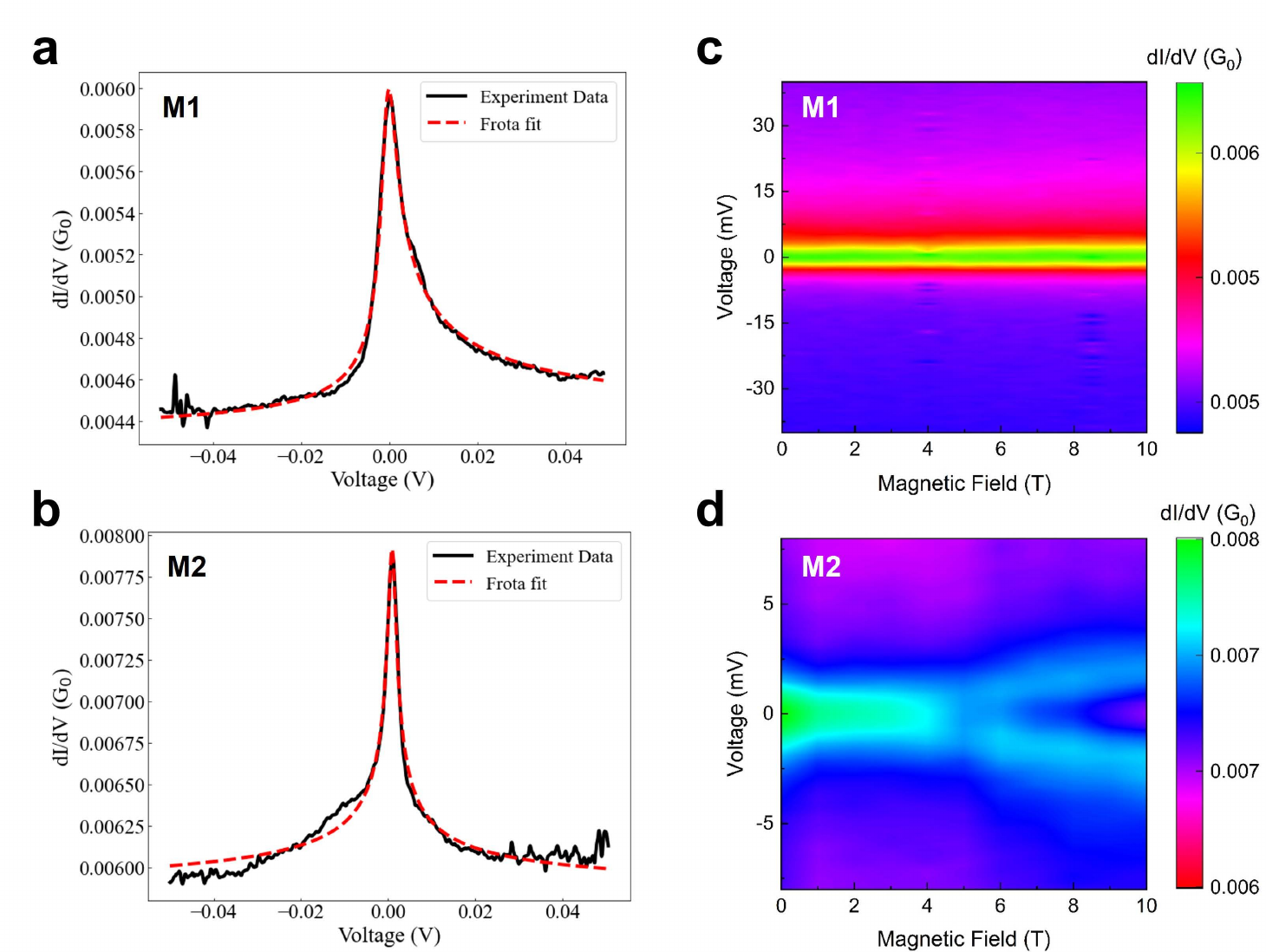}

    \caption{$dI/dV$ analysis of different single-molecule junctions of the Blatter radical. (a)  $dI/dV$ spectra of a type M1 junction showing a broad zero-bias anomaly and fitted using the Frota function. (b) dI/dV spectra of a type M2 junction showing a narrow zero-bias anomaly and fitted using the Frota function. (c) Magnetic field dependence of the $dI/dV$ spectra shown in (a) from 0 -- 10 T at 4.2 K. No splitting of the zero-bias peak is observed here until 10 T.  (d) Magnetic field dependence of the $dI/dV$ spectra shown in (b) from 0 -- 10 T at 4.2 K. A splitting of the zero-bias peak is observed approximately around 7 T. }
     \label{fig:fitting}
\end{figure}

Differential conductance spectroscopy is performed in order to understand the properties of individual Blatter radical molecule junctions in detail.  We discuss the $dI/dV$ characteristics observed in different junctions which are measured using a lock-in amplifier technique 
with 0.1 mV AC amplitude.
These measurements reveal a clear zero-bias anomaly indicating an electronic resonance formed near the Fermi level for about 20\% of single-molecule junctions measured in our study. We have observed that the line width and the line shape of this resonance varies among different junctions under the same measurement conditions. Figures \ref{fig:fitting}a and \ref{fig:fitting}b show examples of two junctions exhibiting different $dI/dV$ spectra. Junction M1 (\ref{fig:fitting}a) shows a broad and asymmetric line shape and junction M2 (\ref{fig:fitting}b) shows a narrower and symmetric resonance. We attribute the origin of these zero-bias anomalies to Kondo resonances due to the presence of an unpaired electron in the current pathway, hence indicating that the open-shell radical nature of the molecule within the two-terminal device is maintained at low temperature. Furthermore we argue that these junctions are monomer junctions, as we will discuss below.\\

For the junction M1, in Fig. \ref{fig:fitting}a the zero-bias peak resembles a Fano resonance line shape as observed in many previous mesoscopic and nanoelectronics devices at low temperature \cite{Madhavan1998, Fano1961, Knaak2017}. This type of spectrum was also previously reported with STM-based studies of the Blatter radical molecule \cite{Low2019a, Patera2019}. A Fano line shape typically occurs when there is an interference between resonant transport channels and non-resonant direct tunneling pathways through the device under study \cite{Zitko2010, Sasaki2009}. Here, these are the transport assisted through the SOMO due to an unpaired spin, interfering with the continuum tunneling channels. The first pathway is responsible for the Kondo resonance while the latter corresponds to the broad conductance background. 
This can be described by the commonly used equation \ref{eq:fano}, 

\begin{equation}
    dI/dV = \alpha\frac{(q+\epsilon)^2}{1+\epsilon^2} + \beta; \epsilon = \frac{eV-E_{\rm K}}{\Gamma_{\rm K}}
    \label{eq:fano}
\end{equation}

\noindent where the Fano parameter, $q$, defines the ratio of transmission amplitudes through the resonant and non-resonant tunneling channels and hence, the peak asymmetry. $E_{\rm K}$ is the energy of the resonant state and $\Gamma_{\rm K}$ gives the halfwidth of the resonance, which is a function of temperature \cite{Nagaoka2002}. Since the experiment is at $k_{\rm B}T\ll \Gamma_{\rm K}$ (verified {\em a posteriori} by our fits), $\Gamma_{\rm K}$ is the zero-temperature halfwidth of the resonance. The constants $\alpha$ and $\beta$ correspond to the peak and background amplitudes, $V$ is the bias voltage, and $e$ is the elementary charge. The above, however, assumes a Lorentzian lineshape for the Kondo resonance. In reality, the Kondo resonance is known from accurate numerical renormalization group (NRG) calculations to have a non-Lorentzian lineshape, and can only be approximated by a Lorentzian at very low energies $|eV|\ll k_{\rm B}T_{\rm K}$ \cite{Jacob2023}. In particular, the Kondo resonance has logarithmic tails decaying as $1/\ln^2(|eV|/k_{\rm B}T_{\rm K})$ at $|eV|\gg k_{\rm B}T_{\rm K}$\cite{Rosch2003}. It is possible to use the essentially exact numerical NRG spectra, together with the Fano effect, to accurately describe Kondo resonance lineshapes in $dI/dV$ \cite{Zitko2011}. Here, we use an approximate analytic approach, the modified
Frota function \cite{Frota1986,Kroha2000}, which simulates well the non-Lorentzian lineshape of the Kondo resonance in the NRG spectra, while including also the Fano effect,
\begin{align} 
dI/dV & = b -c\;{\rm Im} \left[i e^{i\phi}\sqrt{\frac{i\Gamma_{\rm F}}{E-E_{\rm K}+i\Gamma_{\rm F}}}\right]\label{eq:Fano-Frota},
\end{align} 
where $\Gamma_{\rm F}=\Gamma_{\rm K}/2.542$ and $\phi=\phi(q)$ is related to the Fano parameter $q$ (see SI and Fig.~S10). The constants $b$ and $c$ describe the background and Kondo peak amplitude, respectively. Note that both the Fano and modified Frota functions can describe any type of measured resonance, ranging from a symmetric, asymmetric, antisymmetric or even an antiresonance, depending on the Fano parameter $q$ (or $\phi(q)$), see SI. A comparison between the Lorentzian based procedure in Eq.~(\ref{eq:fano}) and the more accurate procedure in Eq.~(\ref{eq:Fano-Frota}) for Kondo resonances is given in the SI, together with a table of all parameters extracted from fitting junctions M1 and M2 to Eq.~(\ref{eq:Fano-Frota}).\\

For the junction shown in Fig. \ref{fig:fitting}a we find the energy width $\Gamma_{\rm K}$ = 4.5 mV. We can extract the Kondo temperature, using the result $k_{\rm B}T_{\rm K}=2 \Gamma_{\rm K}/\pi$ \cite{vanEfferen2024}, and found it to be approximately 34 K. This is in good agreement with the values obtained in previous STM-based studies \cite{Patera2019}. For junction M2, with $\Gamma_{\rm K}= 2.2$ mV, we find $T_{\rm K}$  to be approximately 17 K.\\

 The Kondo conductance and the lineshape depends on the strength of the coupling between the molecule and the electrodes and hence on the bonding sites of the molecule with the electrodes \cite{Appelt2018, Parks2007, Haldane1978}. Hence, we argue that the resonances in cases M1 and M2 originate from different configurations of the molecular junction. Since we have observed both flat and slanted opening traces in our study, the formation of single-molecule junctions with different configurations giving rise to different $T_{\rm K}$ is a likely scenario. For Blatter radicals, it is known that the spin density is mostly centered around the fused benzene ring, benzotriazinyl core and is partially delocalized to one of the phenyl rings attached with the thiomethyl groups \cite{Patera2019, Ji2020}. The third aromatic ring is spin-isolated. Hence, the difference in electronic coupling with both ends of the electrodes can lead to varying coupling strengths and transport pathways. Consequently, the junctions with metal electrodes having a direct interaction with one of the higher spin density sites can display strong zero-bias resonances. We analyzed many individual junctions with conductances between $10^{-2}$ and $10^{-3}$ $G_{0}$ but found no clear correlation with $T_{\rm K}$ and $G$ while 70\% of them show asymmetric resonance with a $T_{\rm K}$ of more than 30 K (type M1). An in-depth theoretical calculation is necessary to distinguish between different Kondo states and junction structures.\\

To further validate the presence of the Kondo resonance and the extracted 
$\Gamma_{\rm K}$ (and $T_{\rm K}=2\Gamma_{\rm K}/\pi k_{\rm B}$) in these junctions, we have recorded the evolution of $dI/dV$ spectra at various magnetic fields from 0 to 10 T at 4.2 K.  A magnetic field typically splits the Kondo resonance due to the Zeeman effect. Figures \ref{fig:fitting}c and \ref{fig:fitting}d show the corresponding spectra for junction M1 and junction M2, respectively. We have observed no splitting or suppression of the zero-bias peak in junction M1 while a clear splitting of the peak is visible for junction M2. Theoretical predictions suggest that this splitting follows the relation, $g\mu_BB_c \approx \Gamma_{\rm K}/2$ \cite{Costi2000, Hsu2021}. For the junction M2, the zero-bias peak shows a clear splitting beyond the splitting field $B_c$ of $\approx 7$ T compared to the theoretical prediction $9.9$ T using the extracted $\Gamma_{\rm K}=2.29$ mV. We do not observe any changes with increasing magnetic field until 10 T for junction M1, in agreement with the estimation from theory (using the extracted $\Gamma_{\rm K}=4.57$ mV) that the splitting in this case should occur only after the $\approx$ 20 T which is above our experimental limit. Further examples for the evolution of $dI/dV$ spectra under magnetic field for both types of junctions is given in the SI. We have found that the experimental field dependence of the $dI/dV$ for all the measured junctions shows a remarkable agreement with the theoretical predictions \cite{Costi2000, Hsu2021}, showing a splitting close to the predicted $B_c$ when this lies in the experimentally accessible field range and no splitting when $B_c$ is predicted to lie outside this range. \\

\subsection*{Negative magnetoresistance and singlet-triplet Kondo effect}
\begin{figure}[t]
    
    \centering
    \includegraphics[width = 0.7\textwidth]{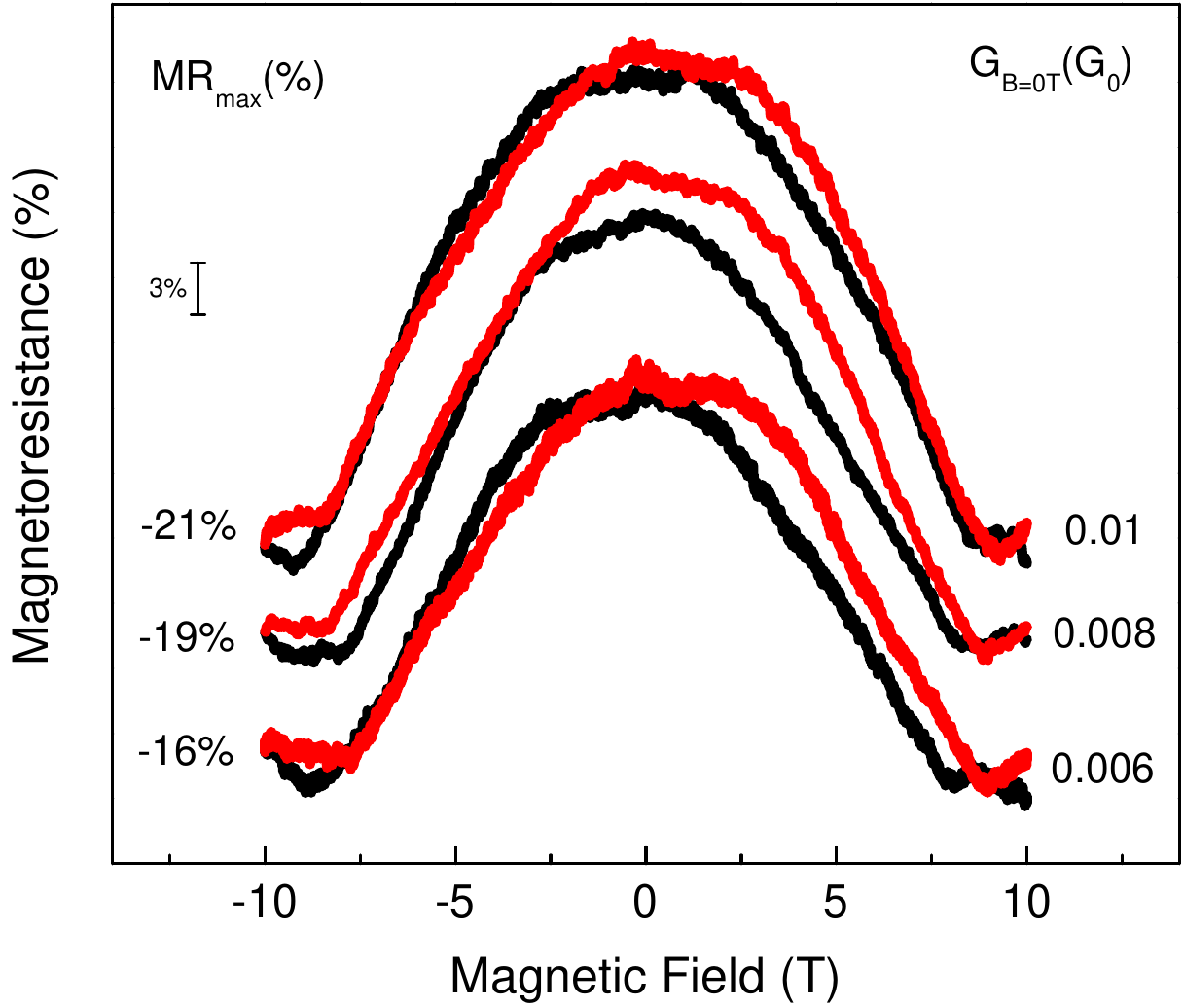}
    
    \caption{Magnetoresistance (MR) measurements of Blatter radical junctions without a zero-bias anomaly  (type D junctions) at different conductances. The magnetic field is swept from $-10$ T to 10 T and back with a sweep rate of 50 mT/min at 4.2 K, and the resistance is measured with an applied bias of 30 mV. The maximum MR\% observed is indicated for each junction. Black curves represent a magnetic field sweep direction from 10 T to $-10$ T and red curves from $-10$ T to 10 T. }
     \label{fig:mr}
\end{figure}

  It was reported earlier that for single-radical molecular junctions without showing any Kondo features, a very pronounced MR can be observed at larger bias \cite{Mitra2022}. MR is defined here as the relative change in resistance of a molecule junction at an applied magnetic field with respect to zero field. In order to study the magnetotransport on molecular junctions which do not show any zero-bias peak in our measurements (denoted as type D junctions), we have studied the evolution of the resistance under magnetic field applied perpendicular to the sample plane from $-10$ to 10 T (red) and 10 to $-10$ T (black) at a rate of 50 mT/min and measured with a constant bias voltage of 30 mV at 4.2 K.  Figure \ref{fig:mr} shows MR curves obtained from three such type D junctions.
We have observed a strong negative MR (i.e. decrease of resistance with increasing field strength) with a maximum change of $\approx$ 21\%. Besides a small hysteresis we observed no clear change in MR amplitude with the sweep direction. MR curves tend to saturate after approximately $\pm 8$ T. We note that in previous studies of radical molecule junctions, this kind of saturation or maxima/minima of the resistance were also observed, but at different field value. We also found that all measured Blatter radical single-molecule junctions show exclusively negative MR in our study. 
In addition, MR measured on type M1 or M2 junctions show only a very small change in resistance of less than 2\%, as given in the SI, hence revealing an anti-correlation between Kondo behavior and strong MR. This behavior was first explored in junctions of the perchlortrityl radical molecule \cite{Mitra2022} and is now confirmed for Blatter junctions. \\
  
  The origin and mechanism of large MR in molecule junctions are still under debate. Various mechanisms have been reported based on the studies of different organic radical molecules with regard to the exact structure of the molecule and anchoring groups. Most of these mechanisms rely heavily on the influence of molecule-electrode interface scattering and changes in the electronic coupling strength \cite{Xie2016a,Shi2017,Hayakawa2016}. In our study, we used methylthioether anchoring units to couple the Blatter radical to the Au electrodes. The previous study on the perchlortrityl radical molecule using the same anchoring groups yielded both positive and negative MR that was assigned to spin-dependent scattering at the molecule-electrode interfaces and resulting in a quantum interference effect \cite{Mitra2022}. In the case of Blatter molecule junctions, we have observed exclusively negative MR and the theoretically reported weak electronic coupling of SOMO and SUMO orbitals with gold electrodes suggests the above MR mechanisms may not be prominent. \\

\begin{figure}[t]

    \begin{minipage}{0.5\textwidth}
    \centering
    \includegraphics[width = 1\textwidth]{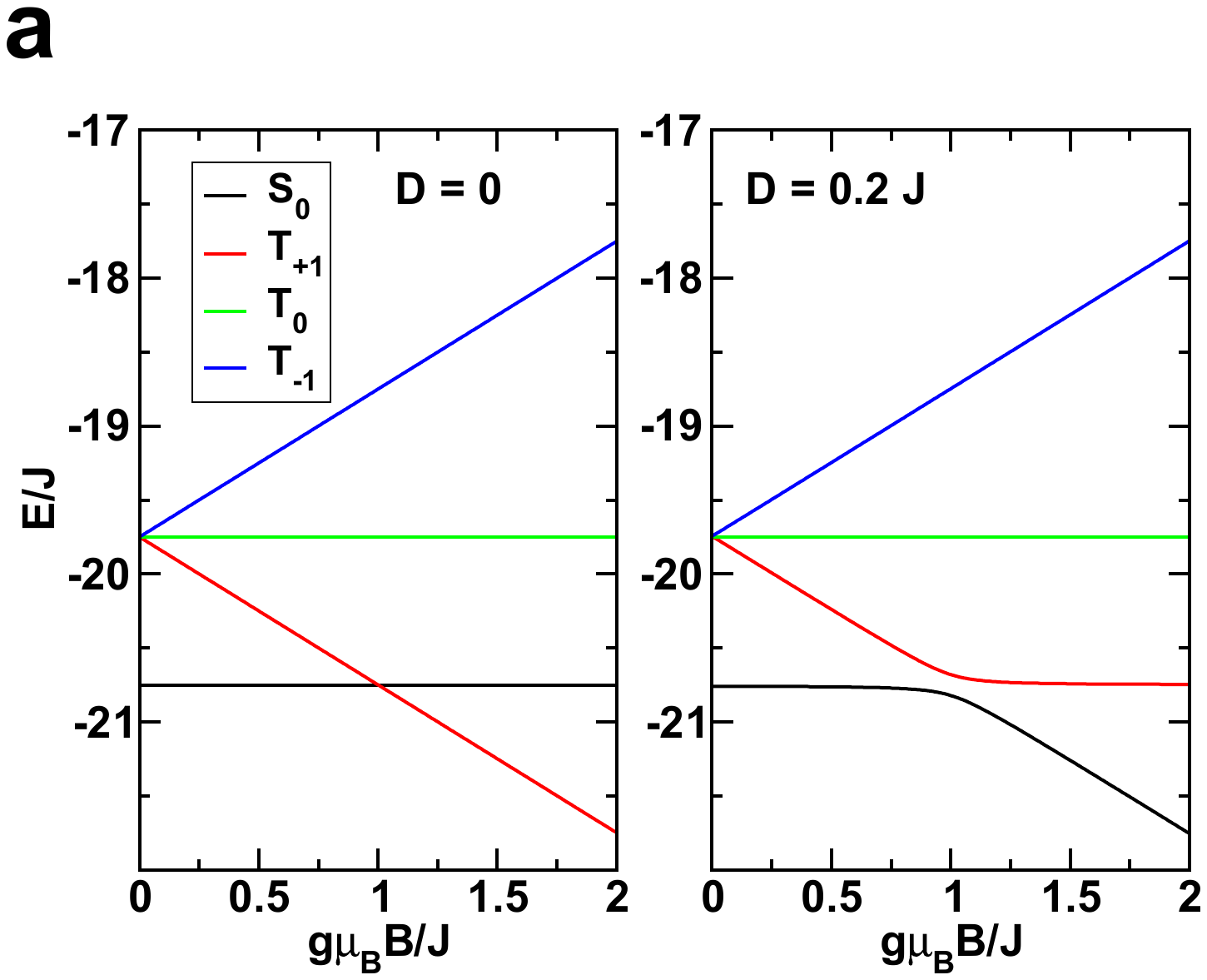}
    \end{minipage}
    \begin{minipage}{0.5\textwidth}
    \centering
    \includegraphics[width = 1\textwidth]{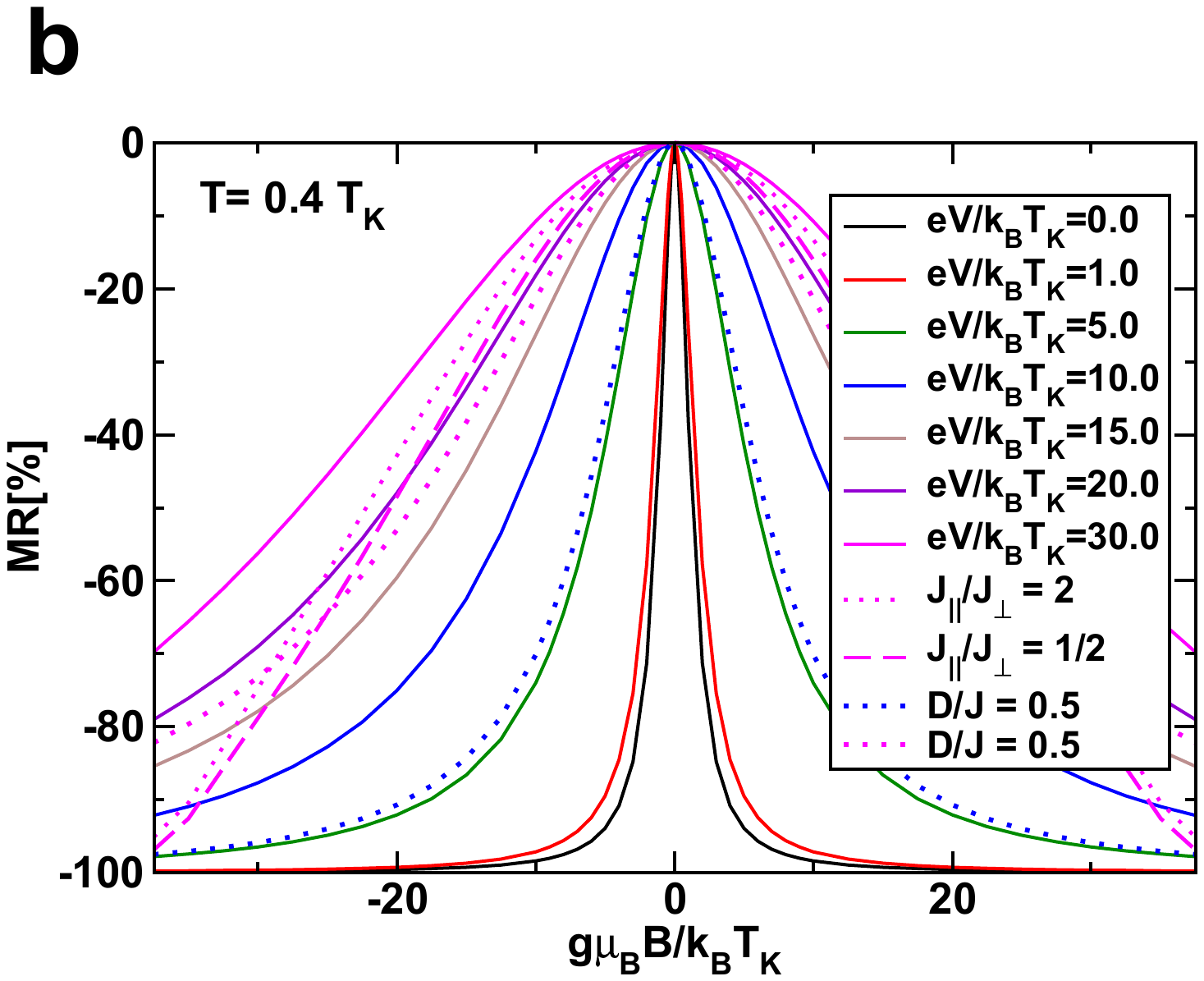}
    \end{minipage}
\caption{(a) Magnetic field dependence of the lowest energy singlet (black, $S_0$) and triplet states ($T_{+1}, T_{-1}, T_0$) of a quantum dot with level energy $\epsilon_0$ = $-20J$ exchange coupled to a second $S = 1/2$ quantum dot with antiferromagnetic coupling with $J = 10$ meV, in the absence, and in the presence of a Dzyaloshinskii-Moriya interaction, $H_D = D\cdot s_d \cross S$. (b) Magnetoresistance at temperature $T = 0.4T_{\rm K}$ = 4 K for different values of the bias voltage $eV/k_{\rm B}T_{\rm K} = 0, 1, \dots, 30$ corresponding (approximately) to $V = 0, 1, \dots, 30$ mV and isotropic exchange J/$k_{\rm B}T_{\rm K}$ = 100 (solid lines). The effect of anisotropic spin exchange is also shown for the largest voltage $V = 30 $mV (magenta solid line): $J_{\parallel}/J_{\perp}$ = 2 with $J_{\parallel}/J_{\perp}$ = 100 (magneta dashed-dotted line) and $J_{\parallel}/J_{\perp}$ = 1/2 with $J_{\perp}/k_{\rm B}T_{\rm K}$ = 100 (magenta dashed line). A small reduction of MR is seen for both longitudinal and transverse anisotropies. The effect of spin anisotropy becomes weaker for smaller voltages. The effect of the Dzyaloshinskii-Moriya interaction is also shown. In order to discern the trends, this is shown for an unusually large $D/J = 0.5$ for both $V = 10$\ mV and $V = 30$\ mV (dotted lines).}
    \label{fig:mr_theory1}
\end{figure}
  
  Hence, we propose a possible explanation of the significant negative MR of type D junctions assuming the junctions actually consist of a double quantum dot system: a Blatter molecule in the junction side-coupled to a second 
  quantum dot, also having a $S = 1/2$, resulting in an antiferromagnetic exchange interaction of strength $J$ between the two (see Eq.~(\ref{eq:Model}) in Theoretical Methods and the SI). The second quantum dot could be, e.g., a gold atom or a cluster with unpaired spin, a molecular fragment, or a second Blatter radical. As mentioned above, the existence of  structures in which a second Blatter radical couples with the first is supported by our quantum chemical calculations (see SI). Furthermore, we found most such pairs to preferentially couple antiferromagnetically, with a coupling constant $J'$ (for a Hamiltonian of the form $H = -2 J' S_a S_b$) in the range of around $-90$ to $-140$ cm$^{-1}$, which translates in our Kondo dimer model notation to antiferromagnetic couplings $J=-2J'$ of order $20-40$ meV, i.e., $200-400$ K.  Examples of the optimized structures of the possible dimer (D) junctions, the resulting transmission functions and the information about energies and spin population are given in the SI. 
  Weak ferromagnetic coupling could also occur for some junctions based on our DFT calculations, but could be pushed to more antiferromagnetic by binding to the electrodes. The latter possibility is also interesting as the D system would form a $S=1$ which would result in an underscreened Kondo effect \cite{Parks2010}.\\

For antiferromagnetic couplings $J$, the double-dot system would exhibit the so called singlet-triplet Kondo effect, which would provide a mechanism for understanding the observed negative MR in the absence of a zero-bias peak in $dI/dV$. 
Organic radical dimers, exhibiting a field-induced singlet-triplet Kondo effect have been investigated previously, however, the MR was not discussed \cite{Zalom2019}. Within this picture, see Fig. \ref{fig:mr_theory1}a, the ground state of the isolated dimer at $B = 0$ is a singlet $S = 0$ state with a triplet $S = 1$ state lying at an energy $J$ above it. Upon coupling the dimer to the leads, the Kondo effect at low energies, temperatures and voltages $(E, k_{\rm B}T, eV \ll J)$ is therefore suppressed and a zero-bias peak in $dI/dV$ is absent for $\abs{eV} < J$. Instead, the differential conductance exhibits a gap of order $2J$, which at high temperatures $k_{\rm B}T>J$ could be smeared out (see Fig.~S7a in SI). Despite the non-magnetic ground state, the system is still strongly affected by a magnetic field via the effect of the latter on the polarizable triplet state: the magnetic field splits the triplet state and decreases the splitting $\delta(B) = E_{S=1,|S_z|=1} - E_{S=0,S_z=0}$ between the lowest component of the triplet state ($S_z = +1$ for $B > 0$ or $S_z = -1$ for $B < 0$) and the singlet state as shown in Figure \ref{fig:mr_theory1}a. This splitting eventually vanishes at a sufficiently large magnetic field $B = B^*$ where $B^*$ is of the order of $J/g\mu_B$ ($B^* = J/g\mu_B$ exactly when the molecule plus side-coupled quantum dot is detached from the leads, but is slightly reduced when it is attached to the leads \cite{Zalom2019}). Fluctuations between the degenerate states at $B = B^*$ lead to a fully developed singlet-triplet Kondo effect via the hybridization of these states to the leads \cite{Pustilnik2000}. In this scenario, increasing $B$ from zero towards $B^*$ induces the singlet-triplet Kondo effect and thereby enhances the conductance $G(B)=dI/dV$, implying that $G(B = 0) - G(B)$ is negative (i.e., positive magnetoconductance), resulting in a negative MR for the field range $-J < g\mu_BB < +J$ (see Fig.~\ref{fig:mr_theory1}b and Fig.~S8). This is opposite to the usual $S = 1/2$ Kondo effect where a magnetic field reduces the conductance (resulting in a positive MR, see Fig.~S9 in the SI).  \\
  
  We now illustrate the generic MR behavior due to the singlet-triplet Kondo effect via numerical renormalization group calculations (see Theoretical Methods and SI). Two energy scales determine the quantitative aspects of the 
  singlet-triplet Kondo effect, the antiferromagnetic dimer coupling $J$ and the Kondo scale $T_{\rm K}$
  in the absence of the side-coupled quantum dot in (\ref{eq:Model}). The former ($J$) is determined by the coupling strength between the two Blatter molecules, while the latter ($T_{\rm K}$) depends on the tunnel couplings of the Blatter molecule in the junction to the electrodes. 
  However, even with reasonable estimates for $J$, the Kondo scale $T_{\rm K}$ remains largely unknown for the experimental system: while theoretically, $T_{\rm K}$ is easily obtained by switching off $J$ in our model, this is clearly not possible for the experimental system (one cannot remove the side-coupled Blatter molecule in order to extract the $T_{\rm K}$ for the monomer).  Thus, we consider a wide range of $J/k_{\rm B}T_{\rm K}$ = 100, 30, and 10 to illustrate the overall qualitative results for the MR. This uncertainty in $T_{\rm K}$
  for the dimer systems, also means that $T/T_{\rm K}$ is unknown. We shall use $T/T_{\rm K}=0.4$ with $T_{\rm K}=10$ K for an experimental temperature $T=4$ K as a typical value, but it should be noted that the MR depends sensitively on the precise value of $T/T_{\rm K}$ (see SI). The bias-voltage dependence of the MR will also be calculated. \\
  
  Figure \ref{fig:mr_theory1}b shows the calculated MR at several dimensionless voltages $eV/k_{B}T_{\rm K}$. For all voltages, the MR exhibits a similar dependence on the magnetic field as in the experiment. It is largest at $V = 0$. Finite $V$ introduces decoherence, leading to a suppression of the MR. Nevertheless, even at $eV/k_{\rm B}T_{\rm K} = 30$, the MR reaches 10\% at $g\mu_BB/k_{\rm B}T_{\rm K} = 10$. The magnitude of the MR is sensitive to the precise $J/k_{\rm B}T_{\rm K}$ and $T/T_{\rm K}$ (see Fig.~S8 for $J/k_{\rm B}T_{\rm K}=30$ and $T/T_{\rm K}=10$). The effect of a spin-exchange anisotropy ($J_\parallel/J_\perp = 2$ and $J_\parallel/J_\perp = 1/2$) is shown for the largest bias voltage ($eV/k_{\rm K}T_{\rm K} = 30$) and found to be small. Its effect is even smaller for smaller bias voltages. In general, for $\abs{eV} < J$ its effect is to enhance the magnitude of the MR. The effect of the Dzyaloshinskii-Moriya (DM) interaction, shown in Fig. \ref{fig:mr_theory1}b for two bias voltages (dotted lines) and a large $D/J=0.5$, is also to enhance the magnitude of the MR, while not affecting its overall $B$-dependence (for smaller $D/J=0.2$, the enhancement is much smaller, see Fig.~S8a). The enhancement in MR $\approx$ $1-G(V, 0)/G(V, B)$ is due to the reduction of $G(V, B)$ in the presence of a finite Dzyaloshinskii vector ($D$), relative to its value for $D = 0$ (since, a finite $D$ enhances the pseudo-magnetic field in the singlet-triplet Kondo effect, thereby reducing $G$ as discussed in \cite{Zalom2019}).  For comparison, we briefly also show the results for the temperature and bias-voltage dependence of a single Kondo-correlated quantum dot described by the Anderson impurity model in the SI, which is also in clear agreement with the previously reported studies \cite{Mitra2022}.

  \subsection*{Conclusions}
 To summarize, we have developed single-molecule junctions of the Blatter radical molecule using a mechanically controlled break junction setup at low temperature in a cryogenic vacuum environment. 
 Our detailed magnetotransport measurements suggest that the open-shell nature of the Blatter junction remains intact in a two-terminal junction.
 Differential-conductance spectroscopy studies on individual contacts reveal two different types of zero-bias anomalies originating from the different configurations of the metal-molecule-metal junction. They are attributed to a Kondo resonance arising from the delocalized unpaired electron orbital situated in the current pathway. In addition, Blatter radical molecular junctions without a zero-bias peak showed strong negative magnetoresistance (MR) with a maximum change of 21\% in resistance measured at high bias voltage. We modeled the origin of these high negative MR towards junctions involving a Blatter molecule with strong correlation with a side-coupled quantum dot configuration which results in a singlet-triplet Kondo effect. The detailed quantum chemical calculations show qualitative agreement between the experiment and theoretical predictions. Our findings reconcile earlier, seemingly contradictory findings on Blatter radical molecular junctions studied in different environments and will open up several new experimental and theoretical avenues to revisit the transport and magnetotransport properties of other radical molecular junctions. They also pave the way for studying various fundamental aspects of the Blatter radical and its derivatives to tailor their properties for spintronic applications.

  \subsection*{Methods}

\subsection* {Experimental Methods}

\textbf{Device Fabrication.} 
The break junction devices were fabricated on a 500 $\mu$m thick, polished insulating Cirlex (Kapton laminate) substrate.  2 $\mu$m polyimide sacrificial layer was spin-coated on the substrate, baked at 130°C for 5 min, then hard-baked at 430°C for 90 min under vacuum. Next, a double-layer resist consisting of methylmetacrylat-methacrylat acida/polymethylmethacrylat (MMA-MAA/PMMA) is spincoated on top and baked in an oven at 170$^{\circ}$C for 30 min. Electron beam lithography was performed on these substrates of size 3 $\times$ 18 mm$^{2}$ in a Zeiss Cross Beam machine at 10 kV acceleration voltage and are developed using 1:3 methyl isobutylketon: isopropylalcohol (MIBK:IPA) solution for 30 s and rinsed with pure IPA. 80 nm of Au were later deposited using the electron beam evaporation. The samples were then etched using anisotropic reactive ion etching in a mixture of oxygen and sulfur hexafluoride at 1 mbar with a power of 50 W for 30 min. This process removes about 600 nm of the polyimide layer below the electrodes.  We prepared a 0.2 mM solution of Blatter molecules in tetrahydrofuran (THF) and then dropcast it onto a freshly prepared MCBJ substrate in nitrogen atmosphere at room temperature and dried for $\approx$ 1 hour. \\

\noindent \textbf{Transport Measurements.} All experimental measurements were carried out using a custom made cryogenic vacuum dipstick equipped with a MCBJ at 4.2 K in a He bath cryostat. The sample is shielded using a copper cap to prevent stray electromagnetic fields. The electrode displacement is calibrated by analyzing the opening curves of a bare Au break junction using the tunneling conductance expression, $G(\delta x) \propto $ exp $ (-2 r \delta x \sqrt{2m\phi}/\hbar)$ where $\phi$ is the work function of Au approximated as 5.0 eV, $m$ is the electron mass, $\delta x $ is the pushing rod movement of the break junction setup, and the reduction ratio $r$ has the relation with electrode displacement, $\delta d = r \delta x$ \cite{Agrait2003}. We found a reduction ratio $r = 3.6 \times 10^{-5}$ which we use to calculate the electrode displacement. At base temperature (4.2 K), sample is repeatedly opened and closed to form atomically sharp Au electrodes and this method allows to form molecular junctions at low temperature \cite{Reichert2003}.  Electronic transport measurements were carried out using a Yokogawa 7651 as a DC voltage source. The wiring is composed of homemade coaxial cables and SMA connectors. Current and voltage across the sample are recorded using Femto DLCPA and Femto DLPVA low noise amplifiers and measured with the help of Agilent 34410A multimeters. $dI/dV$ measurements were carried out using HF2LI Zurich Instruments lock-in amplifiers by applying an AC modulation voltage of 0.1 mV  at 317 Hz. Magnetotransport measurements were performed with an applied magnetic field up to $\pm$ 10 T perpendicular to the sample plane produced from a superconducting magnet inside the liquid-He dewar.  All measurements and devices were remotely controlled using a custom Python program. Control experiments utilizing helium exchange gas for thermalization confirmed that eddy-current heating effects were negligible in our experiments.

\subsection*{Theoretical Methods}  
\textbf{Singlet-triplet model.}
  We model the  singlet-triplet Kondo effect by the Anderson impurity model side-coupled to a second $S=1/2$ quantum dot via an antiferromagnetic exchange interaction of strength $J$. The Hamiltonian is given by
\begin{align}
    H &=H_{\rm dot}+H_{\rm leads}+ H_{\rm tunneling}+H_{\rm J}+H_{\rm B}.\label{eq:Model}
\end{align}
The first term, $H_{\rm dot}=\sum_{\sigma}\varepsilon_{0}d_{\sigma}^{\dagger}d_{\sigma} +Un_{d\uparrow}n_{\downarrow}$, describes the dot Hamiltonian, representing the
$S=1/2$ degrees of freedom of the Blatter molecule,
where $\varepsilon_0$ is the level energy, measured relative to  the Fermi level $E_{\rm F}$, $n_{d\sigma}$ is the occupation number for spin $\sigma={\uparrow,\downarrow}$ electrons on the dot, and $U$ is the local Coulomb repulsion 
on the dot. The second term describes the Hamiltonian of the leads and is given by $H_{\rm leads}=\sum_{k\alpha\sigma}\epsilon_{k}c_{k\alpha\sigma}^{\dagger}c_{k\alpha\sigma}$, where $\alpha={L,R}$ labels the two leads, and $\epsilon_{k\sigma}$ is the kinetic energy of the lead electrons.  The third term, $H_{\rm tunneling}=\sum_{k\alpha\sigma}t_{\alpha}(c_{k\alpha\sigma}^{\dagger}d_{\sigma} +H.c.)$, describes the tunneling of electrons from the leads onto and off the dot with tunneling amplitudes $t_{\alpha}$ and tunneling rates $\Gamma_{\alpha}=\pi \rho_{\rm F}t_{\alpha}^2$, with $\rho_{\rm F}$ the lead electron density of states at the Fermi level. The term $H_{\rm J}=J {\bf s}_{d}\cdot {\bf S}$ describes an antiferromagnetic ($J>0$) coupling between the spin ${\bf s}_{d}$ of the Blatter molecule and the spin ${\bf S}$ of the side-coupled quantum dot. Finally, $H_{\rm B}=-g\mu_{\rm B}BS_{z}^{\rm tot}$ describes a magnetic field acting on the combined spin of the quantum dot and side-coupled quantum dot via the total $z$-component $S_{z}^{\rm tot}=s_{z,d}+S_{z}$. The above is a minimal model for the singlet-triplet Kondo effect \cite{Pustilnik2000,Hofstetter2004,Cornaglia2005,Paaske2006}. It can be generalized to include the effects of anisotropic spin exchange $J {\bf s}_{d}\cdot {\bf S}\rightarrow \frac{J_{\perp}}{2}(s_{d}^{+}S^{-}+s_{d}^{-}S^{+})+J_{\parallel}s_{d,z}S_{z} $, and spin-orbit coupling via a Dzyaloshinskii-Moriya (DM) term $H_{DM}={\bf D}\cdot{\bf s}_{d}\times {\bf S}$. The former (anisotropic spin exchange) does not significantly alter the MR predictions of the isotropic model. The main effect of the latter (DM) is to lift the degeneracy of singlet and triplet states at $B=B^{*}$, resulting in an avoided crossing at this field value \cite{Herzog2010}, see Fig.~\ref{fig:mr_theory1}(a), and therefore a reduced conductance for magnetic fields close to $B=B^{*}$ \cite{Zalom2019}. Hence, the  largest effect of the DM term will be in the magnitude of the MR in the vicinity of $B=B^{*}\approx J$.  For $D/J\ll 1$, the overall qualitative dependence of the MR on field is close to that of the pure singlet-triplet Kondo effect (isotropic $J$ and $D=0$), see Fig.~\ref{fig:mr_theory1}(b).\\

The total dot-lead tunneling rate is denoted by $\Gamma=\Gamma_{\rm L}+\Gamma_{\rm R}$. For a strongly correlated quantum dot we have that $U/\Gamma \gg 1$. We shall use $U/\Gamma=24$ for the calculations, a typical value expected for molecular junctions. For example, estimates of $\Gamma$ from the experiment yielded for the Blatter molecular junction $50-60$ meV. For $U/\Gamma=24$ this implies that $U=1.2-1.4$ eV, which is a reasonable value for a single-molecule device. The relevant energy scales of the above model are $J$ and $k_{\rm B}T_{\rm K}$, where $T_{\rm K}$ is the Kondo temperature for the Anderson impurity model ($J=D=0$). More precisely, our $T_{\rm K}$ is defined as $k_{\rm B}T_{\rm K}= (g\mu_{\rm B})^2/4\chi(0)$, where $\chi(0)$ is the zero-temperature spin susceptibility of the Anderson model. This $T_{\rm K}$ is close to that of the well known Haldane expression $T_{\rm K}^{\rm H}=\sqrt{\Gamma U/2}\exp(\pi\varepsilon_0(\varepsilon_0+U)/2\Gamma U)$ \cite{Haldane1978,Hewson1997}. See SI for other definitions of Kondo scales used in the literature and their interrelationships.
Calculations will be presented for several values of the dimensionless ratio $J/k_{\rm B}T_{\rm K}$ ranging from $1$ to $100$.\\

\noindent \textbf{Numerical renormalization group method.}
The model (\ref{eq:Model}) is diagonalized to obtain its many-body eigenstates and eigenvalues via the numerical renormalization group method \cite{Wilson1975,Bulla2008}. Briefly stated, this approach consists of three steps: (i) the conduction electron kinetic energy of the leads $\sum_{k\alpha\sigma}\epsilon_k c_{k\alpha\sigma}^{\dagger}c_{k\alpha\sigma}$ is logarithmically discretized about the Fermi level $E_{\rm F}$, taken as zero of energy, $\epsilon_k \to \epsilon_{k_n}=\pm D\Lambda^{-n}, n= 0,1,\dots$, with $D$ the half-bandwidth of the leads and $\Lambda>1$ the logarithmic discretization parameter, (ii), the discretized leads are then transformed to a (tridiagonal) linear chain form using a Lanczos procedure, and, (iii), the resulting linear chain form of (\ref{eq:Model}) is iteratively diagonalized on successively lower energy scales to obtain the eigenvalues $E_{p=1,2,\dots}^{n}$ and eigenstates $|p\rangle_n$ on all energy scales $\xi_n =D\Lambda^{-n/2},n=0,1,\dots$. Knowledge of the eigenstates and eigenvalues on all energy scales then allows physical properties such as thermodynamics and Green functions to be calculated. Within this approach, with refinements \cite{Hofstetter2000,Campo2005,Peters2006,Weichselbaum2007,Kugler2022} described in the SI, we calculate the magnetic field, bias voltage and temperature dependence of the differential conductance $G(V,T,B) = dI/dV$ via
\begin{align}
  G(V,T,B)  &= \int_{-\infty}^{+\infty}-\frac{\partial f(E-eV,T)}{\partial E}\, A(E,T,B) dE,\label{eq:gv-main}
\end{align}
where  $f$ is the Fermi function and $A(E,T,B)$ is the spectral function of the impurity level. The general dependence of the spectral function and differential conductance on field, temperature and energy $E=eV$ is shown in Figs.~S5-S7. From $G(V,T,B)$ we extract the magnetoresistance ($MR$) value at bias voltage $V$ and temperature $T$ via
   \begin{align}
    MR[\%] &= 100 \times (R(B)-R(0))/R(0),\\
    & = 100 \times (G(V,T,B=0)-G(V,T,B))/G(V,T,B).\label{eq:MR-main}
  \end{align}

\noindent \textbf{First Principles Calculations.}
All molecular structures were optimized using Kohn-Sham density functional theory (KS-DFT) as implemented in the program package \textsc{Turbomole} 6.6~\cite{bala20}. The B3LYP~\cite{beck93,lee88,vosk80,step94} exchange--correlation functional was used, along with Ahlrichs' def2-TZVP~\cite{weig05} single-particle atom-centered basis set, and Grimme's empirical dispersion corrections (DFT-D3)~\cite{grim10} with Becke--Johnson damping~\cite{grim11}. Convergence criteria of $10^{-7}$ a.u. for the energy in the self-consistent-field (SCF) algorithm and $10^{-4}$ for the gradient in the molecular structure optimizations were applied, in combination with m4 integration grids. 
The coupling constants $J$ are obtained using the Yamaguchi formula \cite{Kitagawa18,yama05} based on the energy difference between an open-shell singlet and open-shell triplet (broken-symmetry) determinant~\cite{nood81}.\\

The conductance of the junctions was evaluated assuming coherent tunneling as the dominant transport mechanism (Landauer regime). The transmissions are modeled via a Green's function approach combined with DFT calculations for zero-bias electronic structures, employing the wide-band limit for the self-energies of the electrodes, as described, e.g., in Refs.\cite{caro71,herr10,herr10a,herr11}. A local density of states (LDOS) of 0.036\,eV$^{-1}$ was used. Transmission functions were evaluated with \textsc{Artaios}~\cite{artaios}, using the Fock and overlap matrices from the KS-DFT electronic structure calculations on the molecular junctions. The value of the Fermi energy is not straightforward to predict in first-principles calculations on molecular junctions, as it will be affected by factors such as the irregular atomistic shapes of the electrodes and the adsorption of non-bridging molecules on the electrodes. We, therefore, evaluate zero-bias conductances for a range of reasonable Fermi energies ($-$5 eV, $-$4.5 eV, $-$4 eV) and check whether our conclusions are robust with respect to this choice. 
For further information please refer to the SI.
  
  \subsection*{Acknowledgements}
 We thank the staff of the nano.lab at the University of Konstanz for technical support. G. M. and E.S. acknowledge funding by the Deutsche Forschungsgemeinschaft  (DFG) through  the Collaborative Research Center SFB 767 ``Controlled Nanosystems'' (funding ID 32152442).  K. S. and C. H. acknowledge funding via the Graduate School 2536 ``Nanohybrid'' of the Deutsche Forschungsgemeinschaft (DFG), funding ID 408076438. M. D. and C. H. acknowledge support by the Cluster of Excellence ``CUI: Advanced Imaging of Matter'' of the Deutsche Forschungsgemeinschaft (DFG) (EXC 2056, funding ID 390715994). T. A. C. gratefully acknowledges computing time on the supercomputer JURECA \cite{JURECA} at Forschungszentrum Jülich under grant no. JIFF23.

 \printbibliography

\end{document}